\PassOptionsToPackage{nomarkers}{endfloat}
\documentclass[useAMS,usenatbib,referee]{biom}
\usepackage[utf8]{inputenc}
\usepackage[T1]{fontenc}
\usepackage{lmodern}
\usepackage{amsmath,amssymb,bm,graphicx}
\usepackage[table]{xcolor}
\usepackage{booktabs,array,tabularx,longtable,multirow,ragged2e}
\usepackage{enumitem}
\setlist{leftmargin=2em,labelwidth=1.5em,labelsep=0.5em}
\usepackage{tikz}
\usetikzlibrary{arrows.meta,positioning,calc,shapes.geometric,fit}
\usepackage[hidelinks,hypertexnames=false]{hyperref}
\makeatletter
\def\ps@titlepage{\let\@oddhead\@empty\let\@evenhead\@empty\let\@oddfoot\@empty\let\@evenfoot\@empty}
\def\ps@draftpages{\let\@oddhead\@empty\let\@evenhead\@empty\def\@oddfoot{\hfil\normalfont\normalsize\thepage\hfil}\let\@evenfoot\@oddfoot}
\makeatother

\definecolor{myblue}{RGB}{28,67,170}
\definecolor{mygreen}{RGB}{32,128,128}
\graphicspath{{plot/}{fig/}}

\newcommand{\InsertResultFigure}[1]{%
  \IfFileExists{plot/#1}{%
    \includegraphics[width=\linewidth,height=0.75\textheight,keepaspectratio]{plot/#1}%
  }{%
    \IfFileExists{fig/#1}{%
      \includegraphics[width=\linewidth,height=0.75\textheight,keepaspectratio]{fig/#1}%
    }{%
      \IfFileExists{#1}{%
        \includegraphics[width=\linewidth,height=0.75\textheight,keepaspectratio]{#1}%
      }{%
        \PackageWarning{manuscript}{Missing figure file: plot/#1}%
        \fbox{\parbox[c][45mm][c]{0.86\linewidth}{%
          \centering\normalfont\small
          Figure file not uploaded.\\[1em]
          \texttt{\detokenize{plot/#1}}\\[1em]
          Upload the original image to display this figure.}}%
      }%
    }%
  }%
}
\newcommand{\PrintReferences}{\bibliography{references}}

\usepackage{comment}
\usepackage{float}

\title[Adaptive Sequential Surveillance]{Adaptive Sequential Surveillance of Infectious Diseases Under Limited Sampling Capacity}

\author{
\shortstack[c]{%
Minghua Xiao, Anne M. Presanis, David S. Robertson,\\
Sofia S. Villar, and Daniela De Angelis
}%
\emailx{daniela.deangelis@mrc-bsu.cam.ac.uk}\\
MRC Biostatistics Unit, School of Clinical Medicine,\\
University of Cambridge, East Forvie Building, Forvie Site,\\
Robinson Way, Cambridge Biomedical Campus,\\
Cambridge CB2 0SR, U.K.
}
\begin{document}
\begin{abstract}
This paper develops a closed-loop adaptive sequential surveillance framework for infectious disease monitoring with limited surveillance resources. In this framework, we use a Bayesian spatio-temporal model to predict disease prevalence at each location, and propose several adaptive site-allocation strategies based on the posterior probabilities that the predictions exceed a predefined threshold. These strategies exploit high-risk locations while also exploring, by random or recency-based sampling or by incorporating the frontiers of emerging outbreaks. To address the unequal inclusion probabilities induced by adaptive sampling, we integrate inverse-probability weighting (IPW) into each model update. A simulation study showed that adaptive surveillance strategies improved outbreak detection relative to uniform sampling and fixed sentinel sampling. Frontier-based strategies improved detection and timeliness relative to risk-proportional allocation. When IPW was used to update the model, some proposed strategies yielded lower inferential bias compared with the standard unweighted update, particularly for strategies that allowed more flexible exploration. Credible-interval coverage varied by strategy. The results identify trade-offs between exploitation of known high-risk locations and exploration of potential outbreak frontiers.

\end{abstract}
\maketitle

\section{Introduction}

Infectious disease surveillance relies on data from a range of sources, including case reports, hospitalisations, positive cases, and wastewater measurements \citep{brainard2023}. These data are used to identify locations with high disease burden, 
how burden changes over time and space, and whether new outbreaks are emerging, supporting the prioritisation of prevention/mitigation strategies \citep{groseclose2017}. 

In practice, many public health agencies make use of fixed surveillance networks: monitoring locations, such as general practices \citep{correa2016}, hospitals and laboratories, are conveniently chosen in advance, and data are routinely collected from these sentinels \citep{cdc2026,bai2017}. This fixed, convenient, design is typically adopted due to a lack of resources to monitor all possible sites \citep{aritaRole2004}. This resource constraint has become very apparent during the COVID-19 pandemic \citep{gonsalves2021, COVID192020}. 

Fixed sentinel designs are very useful for establishing baseline levels and providing consistent information for public health decision-making. However, they have serious limitations. When outbreak detection is the surveillance objective, for instance, 
fixed designs do not capture
spatially localised emerging outbreaks, particularly when such outbreaks occur outside the sites covered by the sentinel network \citep{kasy2020}. Even if a network is representative at the design phase, it may become less informative as outbreaks evolve, shift, or concentrate in particular areas \citep{unkel2012}. Therefore, a practical question is how to decide which locations should be monitored when only a subset of the surveillance network can be sampled and how this design should be adapted as an outbreak progresses. 

To address this challenge, adaptive sequential surveillance approaches have been suggested, e.g.\ \citet{kasy2020, warren2025, fairley2022}, which adjust the sampling design as new data are collected, using information collected in previous surveillance rounds to guide the future sampling decisions. 

A simple example of this type of design is risk-proportional (RP) sampling, in which locations are sampled with probabilities proportional to their predicted outbreak risk \citep {cheung2014}. This type of approach may fail to sufficiently monitor locations currently estimated to be low-risk, where the underlying dynamics could be changing  \citep{fairley2022}. To mitigate this, some allocation rules, originally developed in bandit settings, such as upper confidence bound (UCB) \citep{auer2002} and $\epsilon$-greedy rules \citep{sutton2018}, provide a useful way to balance `exploitation' and `exploration' in adaptive surveillance. Exploitation is a strategy that takes advantage of current knowledge to make a decision, for instance, by directing resources to areas with highest current risk estimates. Exploration, on the other hand, involves acquiring additional information now to make better decisions in future, e.g. to maintain broader surveillance coverage \citep{warren2025}.

Examples of such strategies include
\citet{du2022}, who propose optimization-based allocation under budget constraints, \citet{fairley2022}, focussing on minimizing detection delay through sequential information-driven site selection without accounting for spatial correlation,
and \citet{tan2025} who develop a greedy selection algorithm that integrates both objectives on informativeness and geographical coverage. 
However, in these works sequential decision-making on site allocation that explicitly accounts for both spatial and temporal dependence in disease risk remains under-explored. 

Here we address this gap in the context of outbreak detection, by developing adaptive sampling strategies within a closed-loop sequential framework, where the selection of locations for sampling affects the observed data, and those data then affect the next sampling decision.  Importantly, such sequential adaptive site allocation requires risk predictions, not only for sampled locations, but also for locations that were not observed in the current round. Furthermore, as the observation process is changed by the adaptive site allocation \citep{jalilian2024}, an informative sampling mechanism under which the sampled locations are no longer independent of the spatial process of interest is introduced \citep{hsiao2026}.  If this mechanism is ignored, the estimated risk surface may be distorted: emerging outbreaks are under-sampled and low-risk locations are likely to be under-monitored or missed entirely, while surveillance resources become trapped in a feedback loop in known high-risk areas.

In this work, we contribute to addressing these challenges in several ways. First, we develop and evaluate a Bayesian adaptive sequential surveillance framework, using the probability of testing positive for a  disease (``positivity''), a proxy for disease prevalence, as a measure of risk. We propose several adaptive site-allocation strategies reflecting different surveillance priorities, including prioritisation of locations with high predicted risk, random or recency-based exploration, and incorporation of the frontiers of emerging outbreaks. These strategies are then combined with a Bayesian spatio-temporal risk prediction in a closed-loop process to sample future locations. To address the potential bias in the estimated mean risk across the surveillance network introduced by the adaptive sampling, we adopt an inverse probability weighting (IPW) approach \citep{horvitz1952, hernan2020, hsiao2026}, aiming to reduce systematic distortion of the posterior risk surface. To test the approach, we conduct a simulation study under an epidemic scenario to assess both the detection utility and inferential validity of the proposed strategies and the adaptive sequential surveillance framework. 

The remainder of the paper is organized as follows. Section \ref{sec:framework} describes the proposed adaptive sequential surveillance framework. Section \ref{sec:simulation} details the simulation study design. Section \ref{sec:results}  presents the simulation results. Section \ref{sec:discussion} discusses implications, limitations, and outlines future work.

\section{Adaptive Sequential Surveillance Framework}
\label{sec:framework}

\subsection{Overview }
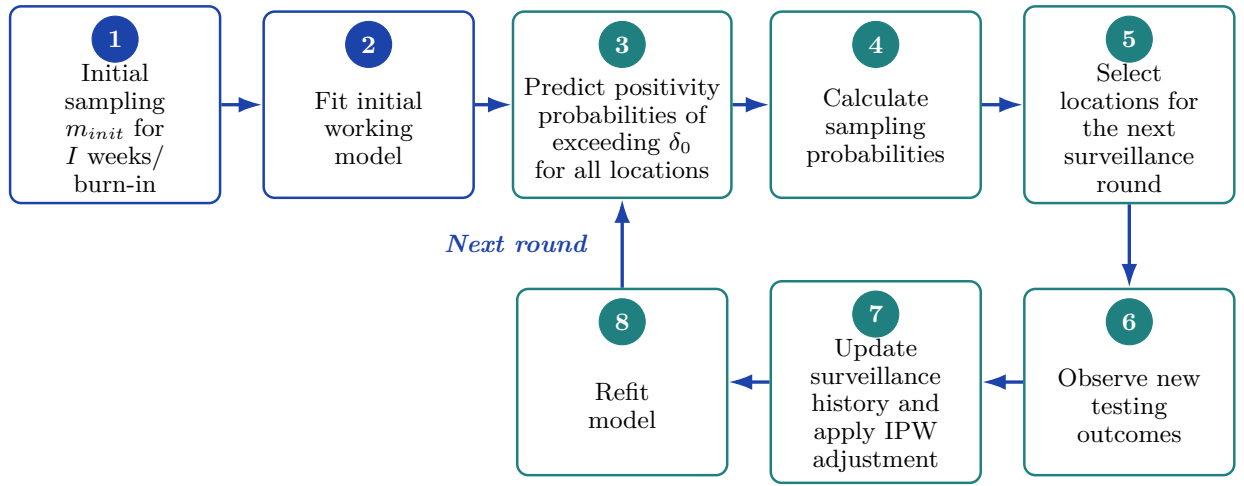
\begin{figure}[p]
\centering
\resizebox{\textwidth}{!}{
\begin{tikzpicture}[
    boxblue/.style={
        rectangle,
        rounded corners=4pt,
        minimum width=2.65cm,
        minimum height=2.35cm,
        align=center,
        draw=myblue,
        line width=0.9pt,
        font=\small
    },
    boxgreen/.style={
        rectangle,
        rounded corners=4pt,
        minimum width=2.65cm,
        minimum height=2.35cm,
        align=center,
        draw=mygreen,
        line width=0.9pt,
        font=\small
    },
    badgeblue/.style={
        shape=circle,
        fill=myblue,
        draw=myblue,
        text=white,
        minimum size=0.58cm,
        inner sep=0pt,
        font=\bfseries\small
    },
    badgegreen/.style={
        shape=circle,
        fill=mygreen,
        draw=mygreen,
        text=white,
        minimum size=0.58cm,
        inner sep=0pt,
        font=\bfseries\small
    },
    arrow/.style={
        -{Latex[length=3mm,width=2mm]},
        line width=1.2pt,
        draw=myblue
    }
]

\node[boxblue]  (n1) at (0,0) {
    \\[0.45cm]
    Initial \\ sampling\\
     $m_{init}$ for\\ $I$ weeks/\\
    burn-in 
};

\node[boxblue]  (n2) at (3.2,0) {
    \\[0.45cm]
    Fit initial\\
    working\\
    model
};

\node[boxgreen] (n3) at (6.4,0) {
    \\[0.45cm]
   Predict positivity \\
     probabilities  of   \\
     exceeding \(\delta_0\)\\
   for all locations
};

\node[boxgreen] (n4) at (9.6,0) {
    \\[0.45cm]
    Calculate\\
     sampling\\
    probabilities
};

\node[boxgreen] (n5) at (12.8,0) {
    \\[0.45cm]
    Select\\
    locations for\\
    the next\\
    surveillance\\
    round
};

\node[boxgreen] (n8) at (6.4,-3.5) {
    \\[0.45cm]
    Refit\\
    model
};

\node[boxgreen] (n7) at (9.6,-3.5) {
    \\[0.45cm]
    Update\\
    surveillance\\
    history and\\
    apply IPW\\
    adjustment
};

\node[boxgreen] (n6) at (12.8,-3.5) {
    \\[0.45cm]
    Observe new\\
    testing\\
    outcomes
};

\node[badgeblue]  at ($(n1.north)+(0,-0.43)$) {1};
\node[badgeblue]  at ($(n2.north)+(0,-0.43)$) {2};
\node[badgegreen] at ($(n3.north)+(0,-0.43)$) {3};
\node[badgegreen] at ($(n4.north)+(0,-0.43)$) {4};
\node[badgegreen] at ($(n5.north)+(0,-0.43)$) {5};
\node[badgegreen] at ($(n6.north)+(0,-0.43)$) {6};
\node[badgegreen] at ($(n7.north)+(0,-0.43)$) {7};
\node[badgegreen] at ($(n8.north)+(0,-0.43)$) {8};

\draw[arrow] (n1.east) -- (n2.west);
\draw[arrow] (n2.east) -- (n3.west);
\draw[arrow] (n3.east) -- (n4.west);
\draw[arrow] (n4.east) -- (n5.west);

\draw[arrow] (n5.south) -- (n6.north);

\draw[arrow] (n6.west) -- (n7.east);
\draw[arrow] (n7.west) -- (n8.east);

\draw[arrow] (n8.north) -- (n3.south);

\node[
    font=\itshape\bfseries\small,
    text=myblue
] at (5.05,-1.75) {Next round};

\end{tikzpicture}
}
\caption{Workflow of the closed-loop adaptive sequential surveillance framework.}
\label{fig:flowchart}
\end{figure}

We consider a surveillance process on $N$ candidate surveillance locations and over a total of $ T$ surveillance weeks.  The proposed adaptive sequential surveillance framework is illustrated in Figure~\ref{fig:flowchart}.  

To initialise the adaptive phase with enough information to fit the Bayesian spatio-temporal model, the process begins with a non-adaptive period  (burn-in) of  $I \ (1 \le I \le T-1)$ weeks, in which a subset of $m_{init}$ locations are uniformly sampled at each week and the positivity is measured as the proportion of tests with a positive result. After the burn-in period, we sample $m$ locations adaptively at each week, denoting the set of locations selected in week $t= I+1, \dots, T$ as  $ {S}_t \subseteq \{1, \ldots, N\}$, and $ | {S}_t|=m $. At the end of the non-adaptive burn-in period, an initial Bayesian spatio-temporal model is fitted to the observed  positivity data from the sampled locations to estimate the risk surface. At the beginning of each subsequent week $t = I+1, \cdots, T$, the  model is fitted to data $\mathcal H_{t-1} $ from a rolling window $\mathcal W_t$ of the previous $q$ weeks, $ \mathcal W_t=\{ s:\max (1,t-q) \leq s \leq t-1 \} $, and used to predict the  probability of the location-specific risk exceeding a threshold \(\delta_0\). This exceedence probability is used to determine the sampling probability $\pi_{it}$ for each candidate location $i$, which are in turn used to choose the sampled set $S_t$. If location $i$ is sampled at week $t$, the sampling indicator $R_{it} = 1$; otherwise, $R_{it} = 0$. Using new testing results from these selected $m$ locations, the model is refitted either with or without the incorporation of the inverse probability adjustment to account for the preferential nature of the adaptive sampling. Notably, different strategies can result in different  surveillance histories, so we distinguish each strategy with the superscript $(a)$.

\subsection{Working Bayesian Spatio-temporal Risk Prediction Model}
\label{sec:framework_model}
A Bayesian spatio-temporal modelling framework is used as the primary working model to  estimate the risk surface across locations and produce one-week ahead risk predictions. The proposed working model combines a time-varying risk level with a local trend that evolves temporally according to a first-order autoregressive [AR(1)] process \citep[Chapter~7]{west1997} and spatially according to the Leroux-type conditional autoregressive (CAR) structure \citep{leroux2000a, west1997}.

Let $n_{it}$ and $Y_{it}$ denote the number of individuals tested  and the number of positive results in location $i$ at week $t $, respectively. We assume that:
\begin{equation}
\begin{matrix}
Y_{it}\mid n_{it},p_{it}\sim \mathrm{Binomial}(n_{it},p_{it})
  \label{eq:predictor}
  \end{matrix}
\end{equation}
where $p_{it}$ is assumed to represent the true disease prevalence.

We model the prevalence with a logistic model 
as: 
\begin{equation}
\begin{matrix}
\mathrm{logit}(p_{it}) = \alpha_0 + u_{it}
  \label{eq:predictor2}
  \end{matrix}
\end{equation}
where $\alpha_0$ is the common intercept and $u_{it}$ is a time-varying risk level on log-odds scale. This risk level is auto-regressively defined in terms of the level $ u_{i,t-1}$ and a local trend $b_{i,t-1}$ at the previous week:
\begin{equation}
\begin{matrix}
  u_{it} =  u_{i,t-1}+\phi b_{i,t-1} + e_{it}

  \label{eq:predictor3}
  \end{matrix}
\end{equation}
where $\phi$ is the autoregressive coefficient of the trend and controls its persistence, with values close to one implying a more persistent trend.  The errors $\boldsymbol{e}_t = (e_{1t}, \cdots, e_{Nt})^T$  
are used to capture the spatial correlation of the risk levels $u_{it}$ for each week $t$. The term $b_{it}$ represents a dynamic local risk trend and is also assumed to follow a spatiotemporal correlation structure:
\begin{equation}
\begin{matrix}

  b_{it} = \phi b_{i,t-1}+z_{it}
  \label{eq:predictor4}
  \end{matrix}
\end{equation}
with $\phi$ defined as above and $\boldsymbol{z}_t = (z_{1t}, \cdots, z_{Nt})^T$ capturing the spatial correlation of the trend for each week $t$. Positive and negative values of $b_{it}$ indicate an upward and downward expected change, respectively. 

The spatial dependence is introduced via a Leroux-type CAR structure \citep{leroux2000a, lee2018a} on both $\boldsymbol{e}_t$ and $\boldsymbol{z}_t$ to borrow information across neighbouring locations while allowing the strength of spatial dependence to be estimated from the data. We assume the following multivariate normal distributions: 
\begin{equation}
    \begin{aligned}
\boldsymbol e_t\mid\boldsymbol\gamma
&\sim \boldsymbol{N}_N\left(\boldsymbol0,
\sigma_{\mathrm{level}}^2\boldsymbol{S}^{-1}(\rho )\right),\\
\boldsymbol z_t\mid\boldsymbol\theta
&\sim \boldsymbol{N}_N\left(\boldsymbol0,
\sigma_{\mathrm{trend}}^2\boldsymbol{S}^{-1}(\rho)\right)
\end{aligned}
\end{equation}
where  $\boldsymbol{\gamma} = (\sigma_{level},\rho) $, $\boldsymbol{\theta} = (\sigma_{trend},\rho) $, and $\sigma_{level}$ and $\sigma_{trend}$ are the standard deviations of the risk level and trend, respectively. The $N\times N$ precision matrix $\boldsymbol{S}(\rho)$ is defined as:
\begin{equation}
    \boldsymbol{S}(\rho) = g(\rho)\cdot[(1-\rho)(\boldsymbol{D}-\boldsymbol{A}) + \rho\boldsymbol{I} ]
\end{equation}
with ${\mathbf A}=( A_{ij})$ denoting the binary adjacency matrix, where $ A_{ij}=1 $ if locations $i$ and $j$ are neighbours and $0$ otherwise; $\boldsymbol{D} = \text{diag}(d_1, \cdots,d_N )$, with $d_i = \sum_j{A_{ij}} $; and scaling factor $g(\rho)$ ensuring the marginal variances of $\boldsymbol{S}(\rho)^{-1}$ have a geometric mean of one. $\boldsymbol{D}-\boldsymbol{A}$ is the graph Laplacian to depict the adjacency structure. $\rho$ controls the borrowing strength of spatial dependence, with larger values indicating weaker spatial correlation and greater independence between neighbouring locations.

Note that this working model is refitted using surveillance history $ \mathcal H_{t}^{(a)} $ at each decision at each week $t$ and produces the strategy-specific posterior predictive distribution of the prevalence $p^{(a)}_{i,t+1}$ at week $t+1$ for all  locations $ i \in \{ 1, \cdots,N\}$. These distributions are then used to calculate the corresponding exceedance probabilities.

\subsection{Adaptive site allocation strategies and sampling probabilities}
\label{sec:framework_strategies}
We use the posterior exceedance probability, $\tilde p^{(a)}_{i,t+1}$, that the prevalence is over a pre-specified threshold \(\delta_0\), defined as:
\begin{equation}
    \tilde p^{(a)}_{i,t+1}= Pr\left(p^{(a)}_{i,t+1} > \delta_0|\mathcal{H}_t^{(a)}\right), i \in\{1,\cdots, N\}
\end{equation} 
as the risk score for each location under each sampling strategy $a$.

From these exceedance probabilities we calculate the strategy-specific sampling probabilities \( \pi^{(a)}_{i,{t+1}}\) for all locations at week \(t+1\), which are used to select $m$ locations for $S_{t+1}^{(a)}$.  

Five sampling strategies are investigated in this paper (as summarised in Table \ref{tab:explore_exploit}): Risk-proportional adaptive surveillance (RP), Risk–recency mixture adaptive surveillance (RP-R),
Average-neighbour frontier adaptive surveillance (ANF), Max-neighbour frontier adaptive surveillance (MNF), and Max-neighbour frontier clipping adaptive surveillance (MNFC).  

\begin{table}[p]
\centering
\caption{Exploration and exploitation components of the five adaptive sampling strategies.}
\label{tab:explore_exploit}

\vspace{8pt}
\renewcommand{\arraystretch}{1.25}
\setlength{\tabcolsep}{4pt}
\fontsize{10}{12}\selectfont

\begin{tabularx}{\textwidth}{@{}
>{\RaggedRight\arraybackslash}p{1.8cm}
>{\RaggedRight\arraybackslash}X
>{\RaggedRight\arraybackslash}X
>{\RaggedRight\arraybackslash}X
@{}}

\toprule
\textbf{Strategy}
& \textbf{Exploitation}
& \textbf{Exploration}
& \textbf{Overall behaviour} \\
\midrule

RP
& Uses the risk-proportional term completely.
& None.
& Prioritises locations with high predicted risk. \\

RP-R
& RP-based, using the risk-proportional allocation with weight $1-\epsilon$.
& Adds a recency-based term that favours locations not sampled recently.
& Risk exploitation with history-aware exploration. \\

ANF
& RP-based, using the risk-proportional allocation with weight $1-\epsilon$.
& Allocates part of the sampling probability according to the Average neighbour frontier score.  
& Risk exploitation with soft frontier-aware exploration for gradual outbreak boundaries. \\

MNF
& RP-based, using the risk-proportional allocation with weight $1-\epsilon$.
&  Allocates part of the sampling probability according to the maximum neighbour frontier score. 
&  Risk exploitation with soft frontier-aware exploration for sharp or localised boundaries. \\

MNFC
&  Uses RP-R as the component of exploitation.
& Adds a probability floor for locations with high maximum neighbouring risk. 
&  Risk exploitation with soft frontier-aware exploration for sharp or localised boundaries via clipping. \\

\bottomrule
\end{tabularx}
\end{table}

\paragraph{\textbf{Risk-proportional adaptive surveillance (RP)}}
This strategy allocates sampling probability in proportion to the predicted risk score. The risk-proportional sampling probability is defined as
\begin{equation}
    \pi^{(RP)}_{i,t+1}
    =
    \frac{\tilde p^{(RP)}_{i,t+1}+\epsilon_0}{\sum_{j=1}^{N}(\tilde p^{(RP)}_{j,t+1}+\epsilon_0)},
    \qquad i=1,\ldots,N,
    \label{eq:risk-prob}
\end{equation}
where $\epsilon_0>0$ is a small constant that ensures a positive allocation score even when the computed exceedance probability is zero.

When disease risks exhibit spatio-temporal dependence, spatial and/or temporal information may help guide exploration. Specifically, monitoring history can help identify locations that have not been sampled recently, while evaluated risk in neighbouring locations may identify areas where outbreaks could emerge or spread. We therefore consider recency-based and frontier-based exploration strategies to incorporate these features into adaptive sampling.

\paragraph{\textbf{Risk--recency mixture adaptive surveillance (RP-R)}}
The RP-R strategy combines the risk-proportional allocation with exploration directed towards locations with high recency scores using the time elapsed since each location was last sampled.

Specifically, the recency score is defined as $g_{i,t}$, representing the number of surveillance rounds since the location $i$ was last sampled before the week $t+1$. The recency score for location $i$ would be higher if it has a longer non-sampled period. Then the corresponding sampling probability is :
\begin{equation}
    \pi^{(RP-R)}_{i,t+1}=(1-\epsilon)\frac{\tilde p^{(RP-R)}_{i,t+1}+\epsilon_0}{\sum_{j=1}^{N}(\tilde p^{(RP-R)}_{j,t+1}+\epsilon_0)}+\epsilon\frac{g_{i,t}+c_g}{\sum_{j=1}^N(g_{j,t}+c_g)},
    \qquad i=1,\ldots,N.
    \label{eq:m2prob}
\end{equation}
where the constant $c_g>0$ ensures a positive probability in the recency component.  If all recency scores are zero, the sampling probability reduces to :
\begin{equation}
    \pi^{(RP-R)}_{i,t+1}=(1-\epsilon)\frac{\tilde p^{(RP-R)}_{i,t+1}+\epsilon_0}{\sum_{j=1}^{N}(\tilde p^{(RP-R)}_{j,t+1}+\epsilon_0)}+\epsilon\frac{1}{N},
    \qquad i=1,\ldots,N.
    \label{eq:m2prob0}
\end{equation}

For frontier-based exploration, we construct ``frontier scores'' by comparing the predicted risk at each location with that of its neighbouring locations. Specifically, the frontier means locations close to high predicted risk locations, but whose own predicted risk has not been as high as those of their neighbours. Sampling these locations helps detect where an outbreak may spread next. We construct two different frontier scores based on either the average or the maximum of neighbours' predicted risk (Figure~\ref{fig:frontier}).

\paragraph{\textbf{Average-neighbour frontier adaptive surveillance (ANF)}}
 The ANF strategy combines the risk-proportional allocation with exploration directed towards locations with high average-neighbour frontier scores, using the average predicted risk (exceedance probability) among neighbouring locations as the frontier baseline.

Specifically, the average predicted exceedance probability among neighbouring locations is calculated as $\tilde   p^{(\mathrm{avg})}_{i,t+1}=\sum_{j=1}^N \tilde A_{ij}\tilde p^{(ANF)}_{j,t+1}$ where $\tilde A_{ij}$ is the row-normal version of $A_{ij}$, and the frontier score is defined as $$ F^{(\mathrm{avg})}_{i,t+1}=\left[\tilde p^{(\mathrm{avg})}_{i,t+1}-\tilde p^{(ANF)}_{i,t+1}\right]_+\cdot \tilde p^{(\mathrm{avg})}_{i,t+1}, $$ where $[x]_+=\max\{x,0\}$. The score increases with the difference between neighbouring and local risk, weighted by the average neighbouring risk, and becomes zero when local risk equals or exceeds neighbouring risk. This allows sampling to target potential outbreak frontiers rather than locations that are already at high risk.

Then the corresponding sampling probability is $$
\pi_{i,t+1}^{ANF}
=
(1-\epsilon)
\frac{\tilde p^{(ANF)}_{i,t+1} + \epsilon_0}
{\sum_{j=1}^{N}\left(\tilde p^{(ANF)}_{j,t+1}+ \epsilon_0\right)}
+
\epsilon
\frac{F_{i,t+1}^{(\mathrm{avg})}}
{\sum_{j=1}^{N}F_{j,t+1}^{(\mathrm{avg})}}
$$ 
If all frontier scores are zero, the sampling probability reduces to :
\begin{equation}
    \pi^{(ANF)}_{i,t+1}=(1-\epsilon)\frac{\tilde p^{(ANF)}_{i,t+1}+\epsilon_0}{\sum_{j=1}^{N}(\tilde p^{(ANF)}_{j,t+1}+\epsilon_0)}+\epsilon\frac{1}{N},
    \qquad i=1,\ldots,N.
    \label{eq:m3prob}
\end{equation}

\paragraph{\textbf{Max-neighbour frontier adaptive surveillance (MNF)}}
MNF is similar to ANF. Instead of using the average predicted risk across neighbouring locations for the frontier score, MNF uses the highest predicted risk among neighbouring locations $\tilde p^{(\max)}_{i,t+1}=\max  A_{ij} \tilde p^{(MNF)}_{j,t+1} $ to calculate the max-neighbour frontier score: $ F^{(\max)}_{i,t+1} =\left[\tilde p^{(\max)}_{i,t+1}-\tilde p^{(MNF)}_{i,t+1}\right]_+\cdot \tilde p^{(\max)}_{i,t+1} $. Then the corresponding sampling probability is $$
\pi_{i,t+1}^{MNF}
=
(1-\epsilon)
\frac{\tilde p^{(MNF)}_{i,t+1}+ \epsilon_0}
{\sum_{j=1}^{N}\left(\tilde p^{(MNF)}_{j,t+1}+ \epsilon_0\right)}
+
\epsilon
\frac{F_{i,t+1}^{(\mathrm{max})}}
{\sum_{j=1}^{N}F_{j,t+1}^{(\mathrm{max})}}.
$$ 

If all frontier scores are zero, the sampling probability reduces to :
\begin{equation}
    \pi^{(MNF)}_{i,t+1}=(1-\epsilon)\frac{\tilde p^{(MNF)}_{i,t+1}+\epsilon_0}{\sum_{j=1}^{N}(\tilde p^{(MNF)}_{j,t+1}+\epsilon_0)}+\epsilon\frac{1}{N},
    \qquad i=1,\ldots,N.
    \label{eq:m4prob}
\end{equation}

\paragraph{\textbf{Max-neighbour frontier clipping adaptive surveillance (MNFC)}}
The MNFC strategy uses a similar frontier score to MNF, but incorporates an additional frontier-clipping-based component, where ``clipping'' refers to setting bounds \citep{JJLee2021}.

A location-specific lower bound on the unnormallised sampling weight is defined as $$ L^{(\mathrm{\max})}_{i,t+1}=c_{\mathrm{floor}} +\rho_F \frac{\{F^{(\mathrm{\max})}_{i,t+1}-\delta_F\}_+} {\sum_{j=1}^N\{F^{(\mathrm{\max})}_{j,t+1}-\delta_F\}_+} $$ where \(c_{\mathrm{floor}}\) is the common lower bound on the sampling weights, ensuring the baseline for all locations.
For locations whose frontier scores exceed the threshold \(\delta_F\), the sampling weight would receive an additional contribution proportional to their excess score. If no frontier score exceeds \(\delta_F\), the denominator is zero and we set \(L_{i,t+1}^{(\max)}=c_{\mathrm{floor}}\) for all \(i\). 

Let $$\pi_{i,t+1}^{\dagger}=(1-\epsilon)\frac{\tilde p^{(MNFC)}_{i,t+1}+\epsilon_0}{\sum_{j=1}^{N}(\tilde p^{(MNFC)}_{j,t+1}+\epsilon_0)}+\epsilon\frac{g_{i,t}+c_g}{\sum_{j=1}^N(g_{j,t}+c_g)},$$ and choose the maximum between it and the frontier floor. Then the unnormalised  probabilities are
\begin{equation*}
    \pi^{\star}_{i,t+1}=
        \max\{\pi_{i,t+1}^{\dagger}, L^{(\max)}_{i,t+1}\}
    \label{eq:m6-unnorm}
\end{equation*}
And the final normalised probabilities are
\begin{equation}
    \pi^{(MNFC)}_{i,t+1}=\frac{\pi^{\star}_{i,t+1}}{\sum_{j=1}^N\pi^{\star}_{j,t+1}}.
    \label{eq:m6prob}
\end{equation}

\subsection{Non-adaptive surveillance strategies}
We also consider two non-adaptive framework benchmarks, Uniform random sampling strategy (URS) and a Sentinel surveillance strategy, to establish a performance baseline against which to compare our adaptive strategies. 

\paragraph{\textbf{Uniform random sampling (URS)}}  This strategy randomly selects locations with the same probability each week, completely ignoring risk predictions, spatial data, or past infection patterns. The sampling probability is defined as:
\begin{equation}
    \pi^{(URS)}_{i,t+1}=\frac{1}{N}, \qquad i=1,\ldots,N.
    \label{eq:m1prob}
\end{equation}
and this strategy provides a non-adaptive benchmark without risk-based optimisation.

 \textbf{Sentinel surveillance} This strategy repeatedly monitors a prespecified subset of locations, ignoring observed positivity history, spatial information, or past infection patterns. Let $\ S_{I+1}$ denote the prespecified set of locations scheduled for routine monitoring in week $I+1$, therefore the sampling set is
\begin{equation}
    S^{(Sentinel)}_{t} = S_{I+1}, \qquad t = I+1,\cdots,T.
    \label{eq:fixed-revisit}
\end{equation}
This strategy is the one currently most often used in practice, e.g. general practices involved in monitoring respiratory infections \citep{correa2016}.

\subsection{Design-weighted pseudo-posterior using inverse inclusion probabilities}
\label{sec:IPW}
Risk-oriented allocation yields unequal sampling coverage across locations. Standard likelihood-based inference for the latent spatio-temporal process may be biased because of the adaptive sampling strategy. To reduce the bias induced by this informative sampling mechanism, we use IPW by incorporating  inverse inclusion probabilities $\Lambda_t:
= \{\lambda_{is}: i=1,s\in \mathcal W_t\}
$ in the design-weighted pseudo-posterior for model (\ref{eq:predictor2}) \citep{savitsky2016a}.  

Define the inclusion probability for location $i$ at time $t$ under strategy $a$ as:
\[\lambda_{i,t}^{(a)}=Pr(i \in S_t^{(a)}|\mathcal{H}_{t-1}^{(a)},a).\] We estimate the inclusion probabilities by repeatedly drawing, without replacement, sets of the $m$ locations via Monte Carlo sampling, i.e., $\hat \lambda_{i,t}^{(a)} = \frac{1}{B}\sum_{b=1}^{B}\mathbf{1}\{i\in S_t^{(a),b}\}$, where \(S^{(a),b}_t\) denotes the sampled location subset obtained in the $b$-th Monte Carlo replicate under strategy $a$, and \(B\) is the number of replications.

Then a truncated inverse-inclusion probability weight \citep{robins2000, hernan2020} is defined as:
\[
\varpi^{0,(a)}_{i,t} =  \min\left\{\frac{\hat \lambda_{i,t}^{-1,(a)}}{\sum_{j \in S_t}\hat \lambda_{j,t}^{-1,(a)}},  \varpi_{max}\right\}
\]
where an upper bound, $\varpi_{max}$, for the weight is applied to cap extremely large weights.
The truncated weights are then normalised within each week to ensure stability:
\[
 \varpi^{(a)}_{i,t} = m \frac{\varpi^{0,(a)}_{i,t}} {\sum_{j\in S^{(a)}_t} \varpi^{0,(a)}_{j,t}}
\]

The weighted binomial log-likelihood for the rolling window $\mathcal{W}_t$ data is then given by:
\begin{equation}
  \ell_{t}^{\mathrm{IPW},(a)}(\boldsymbol \nu) = \sum_{s\in W_t} \sum_{i\in S_s^{(a)}} \varpi_{i,s}^{(a)} \left[ Y_{is}\log p_{is}(\boldsymbol \nu) + (n_{is}-Y_{is}) \log\{1-p_{is}(\boldsymbol \nu)\} \right]
\label{eq:weighted_loglik}
\end{equation}
where \(\boldsymbol \nu\) denotes the collection of parameters in model (\ref{eq:predictor2}). 

The design-weighted pseudo-posterior is: 
\[
\pi_t^{\mathrm{IPW},(a)} (\boldsymbol \nu\mid\mathcal D_t) \propto \exp\{\ell_t^{\mathrm{IPW},(a)}(\boldsymbol \nu)\} \pi_0(\boldsymbol \nu).
\label{eq:pseudo_posterior}
\]
where $ \pi_0(\boldsymbol \nu) $ is the prespecified prior distribution of $\boldsymbol{\nu}$.

\section{Simulation Study Design}
\label{sec:simulation}

The simulation study has three purposes: (1) to examine whether adaptive surveillance improves operational and predictive performance compared with non-adaptive strategies; (2) to quantify the level of bias induced by the risk-oriented sampling and assess whether inverse-probability weighting can mitigate the inferential distortions; (3) to characterize the trade-off between exploitation and exploration across alternative adaptive sampling strategies, and assess how this trade-off changes under different surveillance budgets.

\subsection{Simulation settings}
\label{sec:sim_scenarios}
We conducted the simulations on a $13 \times 13$ spatial grid ($N = 169$ locations) over  $ T = 60$ weeks. The complete underlying surveillance observation dataset for each location at each week was simulated based on the number of individuals tested  \(n_{it}\) and the true prevalences \(p_{it}\) assuming a latent  infectious disease mechanism following susceptible-infected-removed-susceptible (SIRS) model \citep{Keeling2008}, so that all sampling strategies can be evaluated against the same ``true'' latent epidemic process. The SIRS model details are shown in Appendix~\ref{dgp}. The simulated scenario features a single outbreak wave in each location. Heatmaps of true prevalences and outbreak status (prevalence above the risk threshold $\delta_0$) are presented in Appendix Figure~\ref{fig: heatmap}, and the spatial evolution of prevalences over time is illustrated in Appendix Figure~\ref{fig: spatial_positivity}.

We considered four combinations of weekly sampling budget ($m=25$ or $35$ locations) and exploration proportion ($\epsilon=0.1$ or $0.2$). The risk threshold $\delta_0$ for all four scenarios is 0.1 and the burn-in period $I$ is 6 weeks. Each combination was run 50 times with both ``standard unweighted'' and ``IPW weighted'' risk models applied in each replication. The number of Monte Carlo replications $B$ is 2000.

Integrated Nested Laplace Approximation (INLA) is used to provide the posterior summaries of the prevalences for all locations \citep{rue2009}. Specifically, independent priors were assigned to the intercept and hyperparameters: $\alpha_0\sim N\{\text{logit}(0.01),2.5^2\},\ \phi\sim\operatorname{Beta}(4,1),\ \rho\sim\operatorname{Beta}(2,2)$. The intercept prior is centred at a log-odds value corresponding to 1\% prevalence. Exponential priors were assigned to the innovation standard deviations as $\sigma_{\mathrm{level}}\sim\operatorname{Exponential}(r_{\mathrm{level}}),\ \sigma_{\mathrm{trend }}\sim\operatorname{Exponential}(r_{\mathrm{trend }})$ using the rate parameterization, with
$ r_{\mathrm{level}}=-\log(0.05)/2,\ r_{\mathrm{trend }}=-\log(0.05)/0.5$.
These choices imply $ P(\sigma_{\mathrm{level}}>2)=0.05 $ and
$P(\sigma_{\mathrm{trend }}>0.5)=0.05$, and favour smaller week-to-week random changes in the risk level and local trend.
\subsection{Evaluation Metrics}
\label{sec:sim_metrics}
We evaluated the performance of each surveillance strategy from two perspectives, operational utility and inferential validity. Recall that $I$ denotes the number of weeks in the burn-in period, and let 
\[
\mathcal{M}=\{(i,t): i=1,\ldots,N,\ t=I+1,\ldots,T\}
\]
be the set of location--week observations in the adaptive surveillance period. Let $p^{\mathrm{true}}_{it}$ be the true latent prevalence, and 
define the $k$th true outbreak indicator as $ H_{i_kt}=\mathbf{1}\{p^{\mathrm{true}}_{it}>\delta_0\} $ and the corresponding outbreak episode as $[a_{i_k},b_{i_k}]$.

We further define the observed outbreak indicator as $C_{it} = R_{it}\textbf{1}[y_{it}/{n_{it}}>\delta_0]$, where $\delta_0$ denotes the outbreak threshold; the $k$th observed detection time is defined as  $d_{i_k}=\min\{t_k\in[a_{i_k},b_{i_k}]: C_{it_k}=1\} $. If the true outbreak is undetected, e.g. no observed outbreak detection occurred within its interval $ [a_{i_k},b_{i_k}] $, $d_{i_k} = \infty $. The observed outbreak detection delay is defined as $D_{i_k} = d_{i_k}-a_{i_k}$.

 For metrics on detection timeliness, we only include the true outbreaks beginning within $( I,T-1]$ weeks, denoted as $\mathcal{K}= \{i: I < a_{i_k} \le T-1\}$, and $|\mathcal{K}|=K$.

\textbf{Operational utility metrics}
\begin{itemize}[leftmargin=2em]

  \item[\textbf{i.}]\textbf{Outbreak Detection Rate (ODR) }: measures the probability of detecting at least one outbreak event among the true latent outbreaks.
  \begin{equation*}
            \mathrm{Outbreak\ Detection\ Rate} = \frac{1}{K}\sum_{i_k \in \mathcal{K}}\textbf{1}(D_{i_k} <\infty).
        \end{equation*}

  \item[\textbf{ii.}]\textbf{Rate of detection  within $k$ weeks}: measures the proportion of outbreaks that are sampled and therefore detected within $k$ weeks of their onset. 
  
\begin{equation*}
\begin{aligned}
    \mathrm{Detection \ within \ 1 \ week} = \frac{1}{K}\sum_{i_k \in \mathcal{K}}\textbf{1}(D_{i_k} =0),\\ 
        \mathrm{Detection \ within \ 2 \ weeks} = \frac{1}{K}\sum_{i_k \in \mathcal{K}}\textbf{1}(D_{i_k} \le 1).
\end{aligned}      
 \end{equation*}
   
 \item[\textbf{iii.}]\textbf{Mean detection delay}: defined as the average number of weeks from the start of an outbreak to its first detection through sampling. The average was calculated only for outbreaks detected before they ended.
    \begin{equation*}
        \mathrm{Mean\ Delay} = \frac{\sum_{i_k:D_{i_k}< \infty}D_{i_k}}{\sum_{i_k \in  \mathcal{K}}\textbf{1}(D_{i_k} < \infty)},\\ 
    \end{equation*}
 \end{itemize}

\textbf{Inferential validity metrics}
Let $\hat p_{it}$ denote the posterior mean of $p^{\mathrm{true}}_{it}$. Let $\mathcal{G}$ denote the set of location--week observations used for inferential evaluation, with $|\mathcal{G}|=n_{\mathcal{G}}$. In the main analysis, $\mathcal{G}$ includes all location--week observations after the burn-in period.

\begin{itemize}[leftmargin=2em]
    \item[\textbf{i.}]\textbf{Bias}: measures the average difference between the estimated and true risks.
     \begin{equation*}
        \mathrm{Bias} = \frac{1}{n_{\mathcal{G}}} \sum_{(i,t)\in\mathcal{G}} \left(\hat p_{it}-p^{\mathrm{true}}_{it}\right).
     \end{equation*}
    
    \item[\textbf{ii.}]\textbf{Mean absolute error}: measures the average absolute difference between the estimated and true risks.
     \begin{equation*}
        \mathrm{MAE} = \frac{1}{n_{\mathcal{G}}} \sum_{(i,t)\in\mathcal{G}} \left|\hat p_{it}-p^{\mathrm{true}}_{it}\right|.
     \end{equation*}

    \item[\textbf{iii.}]\textbf{Root mean squared error RMSE}: measures the square root of the average squared estimation error and gives greater weight to larger errors.
    \begin{equation*}
        \mathrm{RMSE} = \sqrt{ \frac{1}{n_{\mathcal{G}}} \sum_{(i,t)\in\mathcal{G}} \left(\hat p_{it}-p^{\mathrm{true}}_{it}\right)^2 }.
    \end{equation*}

  \item[\textbf{iv.}]\textbf{Coverage of the 95\% credible interval}:  measures the proportion of true risks that fall within the posterior 95\% credible intervals. Let $\mathrm{CI}^{L}_{it}$ and $\mathrm{CI}^{U}_{it}$ denote the lower and upper bounds of the posterior 95\% credible interval for $\hat p_{it}$. Then
    \begin{equation*}
        \mathrm{Coverage}_{95} = \frac{1}{n_{\mathcal{G}}}\sum_{(i,t)\in\mathcal{G}} \mathbf{1} \left\{\mathrm{CI}^{L}_{it}  \leq p^{\mathrm{true}}_{it} \leq \mathrm{CI}^{U}_{it} \right\}.
    \end{equation*}

   \item[\textbf{v.}]\textbf{Brier score} \citep{brier1950verification}: measures predictive ability through the mean squared error of the predicted exceedence probability $\tilde p_{it}$ compared to the true outbreak indicator $H_{it}$.
\begin{equation*}
    \mathrm{Brier} = \frac{1}{n_{\mathcal{G}}} \sum_{(i,t)\in\mathcal{G}} \left( \tilde p_{it}-H_{it} \right)^2.
\end{equation*}

\end{itemize}

\section{Simulation Results}
\label{sec:results}

\subsection{Comparison of Operational Utility Between Adaptive Sequential Surveillance strategies and the non-adaptive Benchmarks}
\begin{figure}[p]
\centering
\InsertResultFigure{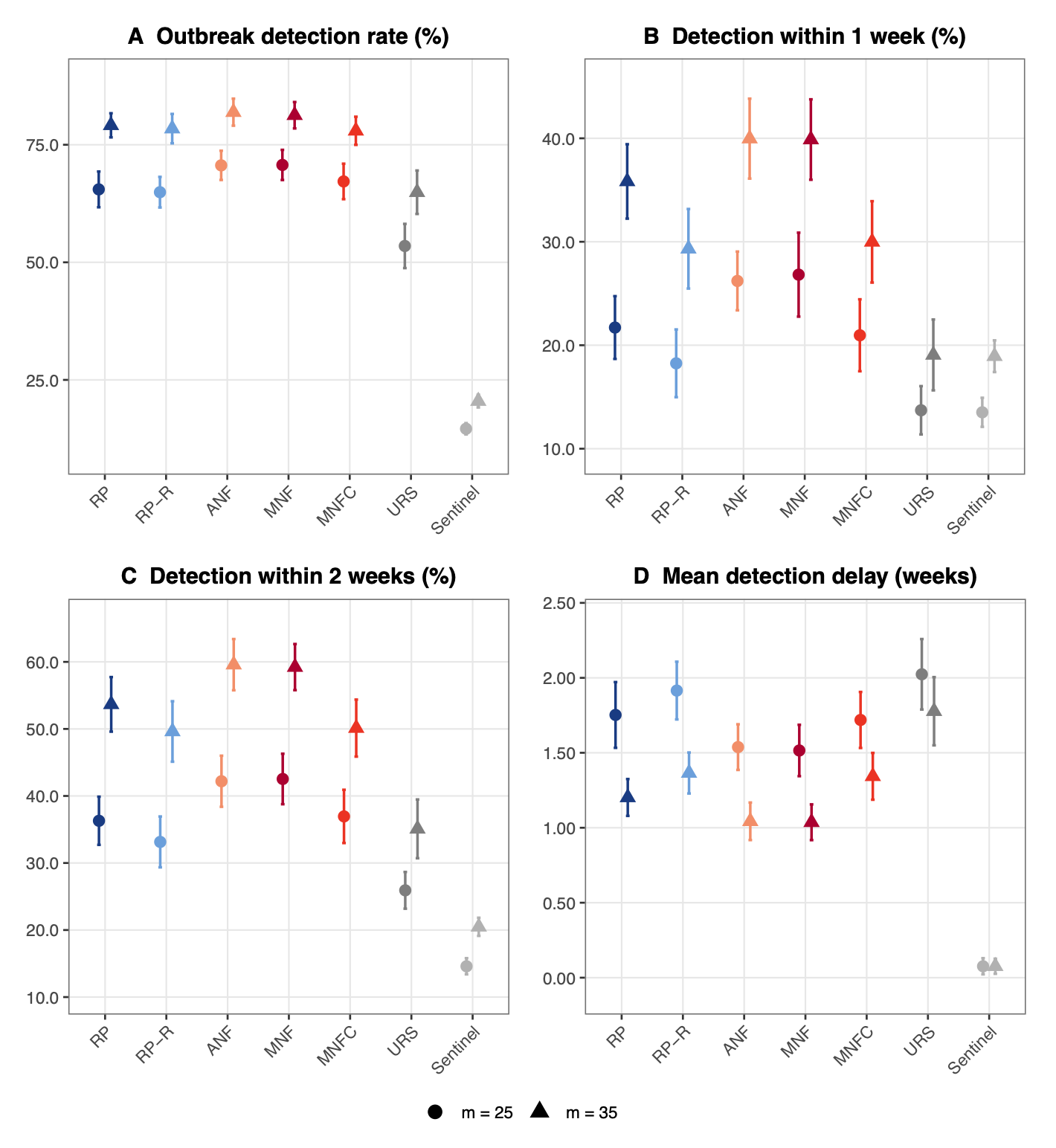}
\caption{\textbf{Operational performance of adaptive sequential surveillance strategies and non-adaptive benchmarks under weekly sampling budgets of 25 and 35 sites in the simulated single-wave scenario.} \(\epsilon = 0.1\).  Panels show (A) outbreak detection rate, (B) detection within one week, (C) detection within two weeks, and (D) mean detection delay. Detection within one week refers to detection in the outbreak onset week; detection within two weeks includes the onset week and the following week. Mean detection delay is calculated only among detected outbreak events.}
\label{fig:detection_35_25}
\end{figure}

Figure~\ref{fig:detection_35_25} and Appendix Tables~\ref{tab:budget_detection} summarise the operational utility  results across different sampling budgets (25, 35) for the simulated single wave scenario. 
All metrics are reported as mean (SD) across simulated replications.

Across both sampling budgets, the adaptive sequential surveillance strategies showed higher ODRs than uniform random sampling (URS) and sentinel sampling (see Figure~\ref{fig:detection_35_25} A and Table~\ref{tab:budget_detection}). When using a weekly sampling budget of $m=35$, the adaptive strategies achieved ODRs of at least 78.0\%, with ANF achieving the highest (81.9\%), followed by MNF with 81.3\%. The same pattern was observed with $m=25$ locations, which showed a small decrease on ODR relative to $m=35$ locations. 

Meanwhile, ANF and MNF also showed the best timely detection performance (see Figure~\ref{fig:detection_35_25} B, C, D and Table~\ref{tab:budget_detection}). When using a weekly sampling budget of $m=35$, the detection rates within one week were 40.0\% and 39.9\%, respectively, and those within two weeks were 59.6\% and 59.2\%. The corresponding rates were 19.1\% and 35.1\% for URS, and 18.9\% and 20.5\% for sentinel sampling. Both ANF and MNF also showed shorter mean detection delays, with 1.043 weeks for ANF and 1.037 weeks for MNF, compared with 1.777 weeks for URS. Sentinel sampling showed a low ODR (20.5\%) but a short mean detection delay (0.076 weeks). Continuous monitoring enabled rapid detection at sentinel sites, while missed outbreaks elsewhere were excluded from the delay calculation. Its short conditional delay therefore did not indicate better detection performance across all outbreak events.

\subsection{Predictive accuracy and interval coverage of the proposed Adaptive Sequential Surveillance strategies and the Non-adaptive Benchmarks}
\begin{figure}[p]
\centering
\InsertResultFigure{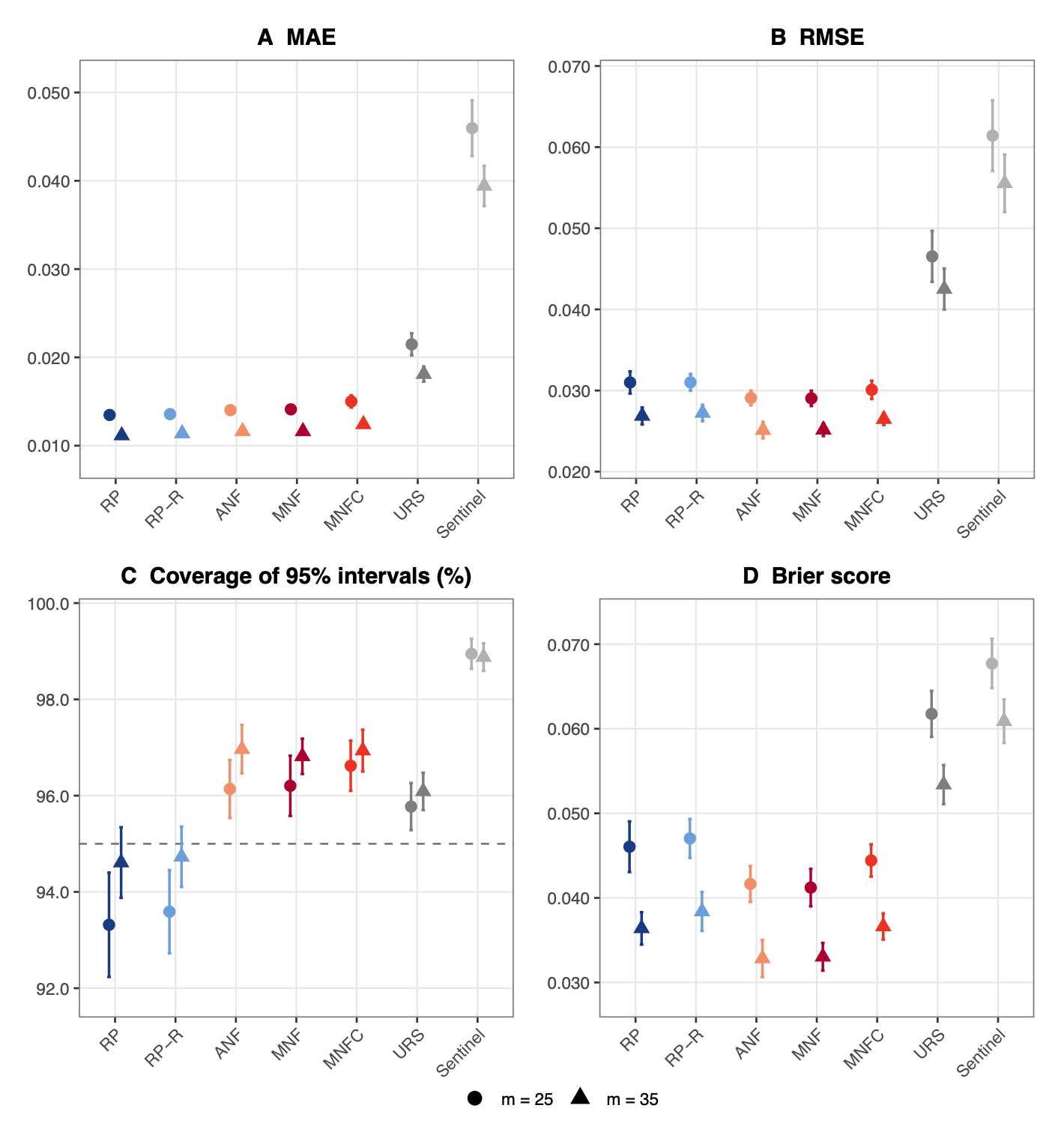}
\caption{\textbf{Predictive performance of adaptive sequential surveillance strategies and non-adaptive benchmarks under weekly sampling budgets of 25 and 35 sites in the simulated single-wave scenario.} \(\epsilon = 0.1\). Panels show (A) mean absolute error (MAE), (B) root mean squared error (RMSE), (C) empirical coverage of posterior 95\% credible intervals, and (D) Brier score. Lower MAE, RMSE, and Brier scores indicate better predictive performance. The dashed horizontal line in panel (C) indicates the nominal coverage of 95\%. }
\label{fig:estimation_35_25}
\end{figure}

 Figure~\ref{fig:estimation_35_25}  and Appendix Table~\ref{tab:budget_estimation} summarise predictive accuracy and interval coverage results under weekly sampling budgets of 25 and 35 sites in the simulated single-wave scenario. Lower MAE, RMSE, and Brier scores indicate better predictive performance. Bias closer to zero indicates less systematic prediction error, whereas coverage closer to the nominal level of 0.95 indicates better interval calibration.

The adaptive strategies achieved lower prediction errors and Brier scores than the benchmarks. With a weekly sampling budget of $m=35$ locations, MAE ranged from 0.011 to 0.012 and RMSE from 0.025 to 0.027 across the adaptive strategies, compared with 0.018 and 0.043 for URS, and 0.039 and 0.056 for sentinel sampling. ANF and MNF both achieved RMSE values of 0.025 and Brier scores of 0.033, compared with Brier scores of 0.053 for URS and 0.061 for sentinel sampling. Coverage of the 95\% credible intervals was between 94.6\% and 97.0\% for the adaptive strategies. It was slightly below 95\% for RP and RP-R. Coverage was 96.1\% for URS and 98.9\% for sentinel sampling. The overall predictive accuracy advantages of adaptive strategies over the benchmarks persisted under the smaller weekly budget of $m=25$ locations. However, compared with $m=35$, MAE, RMSE, and Brier scores were higher, and interval coverage was generally lower.

\subsection{Impact of inverse-probability weighting in the proposed adjusted surveillance framework}
\begin{figure}[p]
\centering
\InsertResultFigure{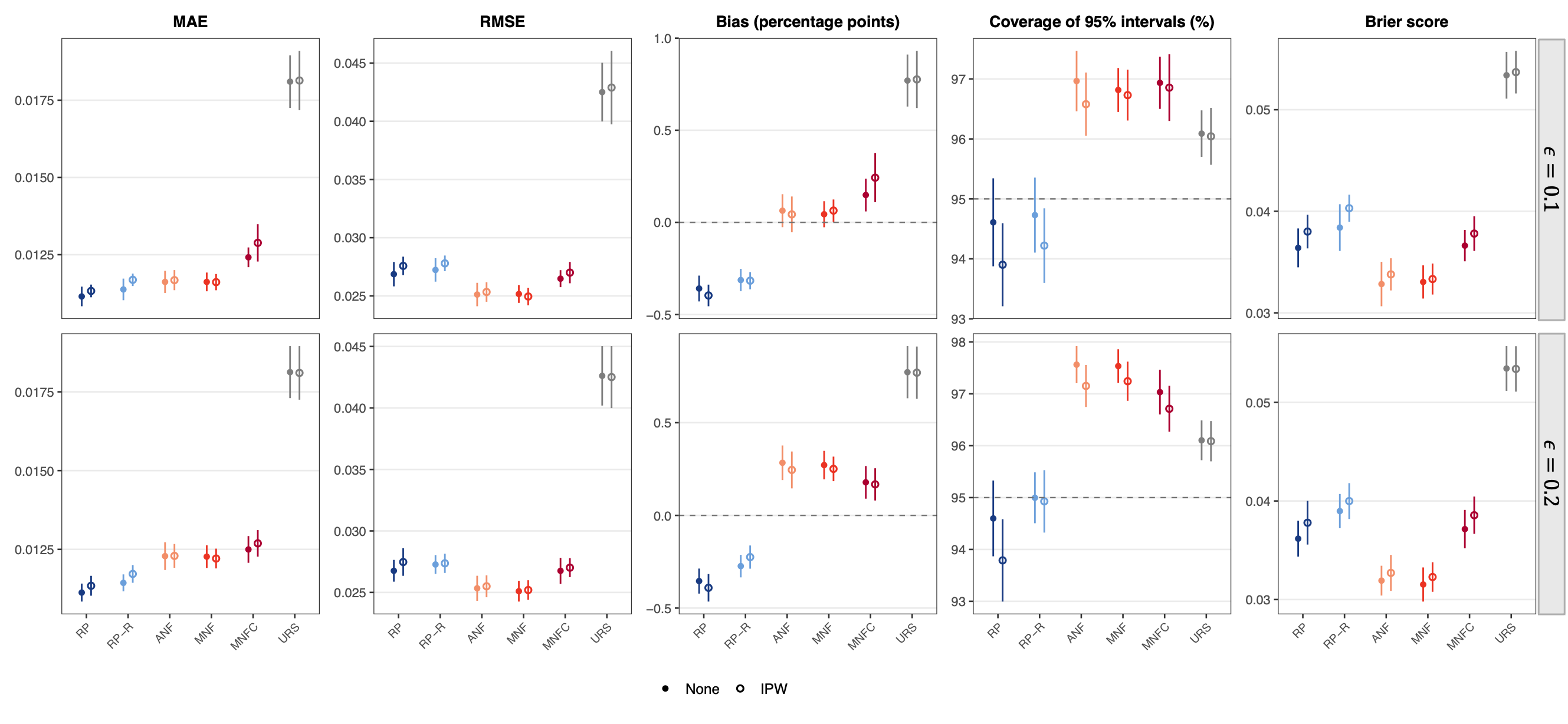}
\caption{\textbf{Overall predictive performance of the adaptive surveillance framework with and without inverse-probability weighting (IPW). }  Budget = 35. MAE, RMSE, bias, coverage of the 95\% credible intervals, and Brier score are shown for \(\epsilon=0.1\) (top row) and \(\epsilon=0.2\) (bottom row).}
\label{fig:ipw_inference}
\end{figure}

Figure~\ref{fig:ipw_inference}  and Appendix Table~\ref{tab:ipw_comparison} compare the predictive performance of the unweighted and inverse-probability weighted (IPW) models under exploration proportions of $\epsilon=0.1$ and $\epsilon=0.2$, with a weekly sampling budget of 35 sites in the simulated single-wave scenario. Overall, the unweighted model already exhibited small biases across the proposed adaptive strategies, ranging from $-0.0036$ to $0.0015$ at \(\epsilon=0.1\) and from $-0.0035$ to $0.0028$ at \(\epsilon=0.2\). Incorporating IPW resulted in modest reductions in absolute bias for some strategies, particularly at the higher exploration proportion. For example, at \(\epsilon=0.2\), the bias of RP-R moved closer to zero, from $-0.0027$ to $-0.0022$, while the absolute bias of ANF and MNF decreased from $0.0028$ to $0.0025$ and from $0.0027$ to $0.0025$, respectively. The effects of IPW on predictive accuracy varied across strategies. For MNF, IPW reduced overall RMSE from 0.0252 to 0.0249 at \(\epsilon=0.1\), with a slight increase at \(\epsilon=0.2\) from 0.0251 to 0.0252.

\begin{figure}[p]
\centering
\InsertResultFigure{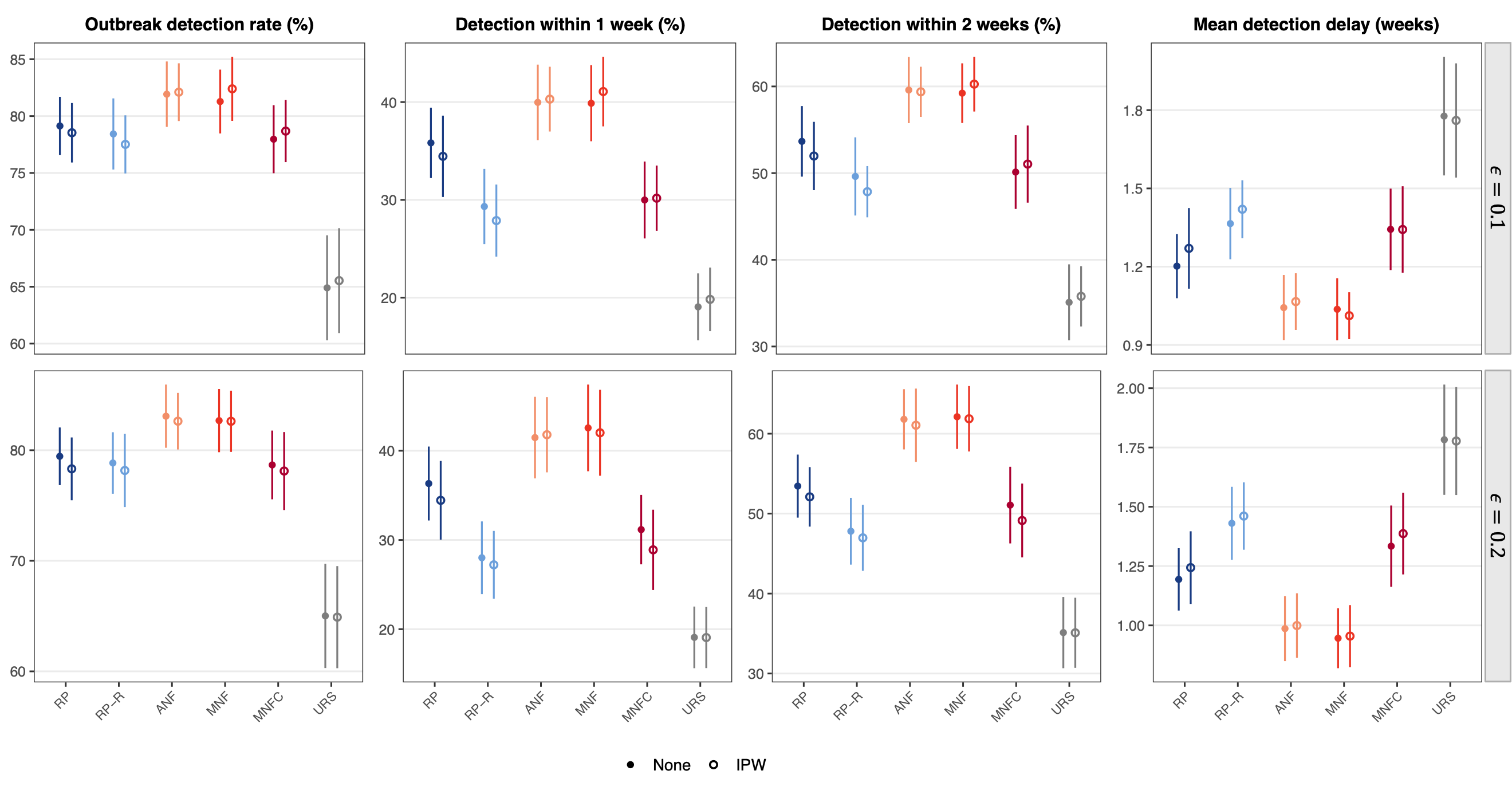}
\caption{\textbf{Operational performance of the adaptive surveillance framework with and without inverse-probability weighting (IPW).} Budget = 35. Outbreak detection rate, detection within one week, detection within two weeks, and mean detection delay are shown under exploration proportions of \(\epsilon=0.1\) (top row) and \(\epsilon=0.2\) (bottom row).}
\label{fig:ipw_detection}
\end{figure}

The incorporation of IPW generally preserved the operational performance of the adaptive strategies at both exploration levels \(\epsilon=0.1, \epsilon=0.2\) (see Figure~\ref{fig:ipw_detection}). For example, for MNF at \(\epsilon=0.1\), ODRs were 81.3\% and 82.4\%, detection within two weeks was 59.2\% and 60.3\%, and mean delay was 1.037 and 1.012 weeks, without and with IPW respectively. Corresponding differences remained small at \(\epsilon=0.2\), with ODRs of 82.7\%  and 82.6\% without and with IPW, two-week detection rates of 62.1\% and 61.9\%, and mean delay of 0.946 and 0.955, respectively. Similar small differences were observed for the other adaptive strategies.

\section{Discussion}
\label{sec:discussion}
We have developed a closed-loop adaptive sequential surveillance framework for infectious disease monitoring with limited surveillance resources. Simulation results showed that this framework substantially improved operational performance, as measured by outbreak detection and detection timeliness, compared with the non-adaptive benchmarks. Overall, the proposed frontier-based strategies (ANF, MNF) were superior for most operational and inferential metrics considered even when the test budget is very limited. 

Frontier-based strategies, including ANF, MNF, and MNFC, showed better performance in outbreak detection and detection timeliness because they are not only oriented by the predicted risks, but also utilize the spatial dependence information to guide the exploration of more locations with potential risk. 
Even under the quite limited resource constraints, with only 25 or 35 sampled out of 169 locations (15\% and 20\% respectively), these proposed strategies showed considerable benefits. This finding has practical value because although incomplete coverage is common in real-world surveillance systems, sentinel systems sometimes have larger coverage than the scenarios we have considered.  For example, the Oxford–RCGP Research and Surveillance Centre network includes almost a third of general practices in England \citep{gu2024}. Our proposed strategies have the potential to improve the surveillance efficiency of such networks.

The exploration proportion plays an important role in balancing exploitation and exploration in adaptive surveillance, resulting in a potential trade-off between operational and inferential performance. A lower proportion of exploration means that resource allocation is more risk-oriented, potentially improving immediate detection of outbreaks. In contrast, a higher proportion of exploration would preferentially discover more locations that might be overlooked, maintaining broader surveillance coverage, and identifying emerging risks. Our simulation results suggest that the effects of increasing the exploration from $\epsilon=0.1$ to $\epsilon=0.2$ varied between strategies and performance measures, showing that the balance between exploitation and exploration depends on the specific surveillance objectives.
 
IPW adjustment improved point estimation of the risk surface, for example under ANF, which allows more exploration, as more locations with small probabilities become part of the sampling subset and improve representativeness of the sample. Similar results were found when more locations were sampled but inclusion probabilities were less extreme.
In contrast, when the adaptive design became highly concentrated with a smaller exploration proportion, some locations had very small inclusion probabilities, resulting in more unstable weights, which limited the potential benefit of IPW \citep{horvitz1952, hernan2020}. This is consistent with other studies where IPW was used to correct the bias introduced by preferential sampling \citep{ hsiao2026}. 

Importantly, although IPW improved point estimates, it did not consistently improve credible interval coverage.
One possible reason is that the credible intervals derived from the weighted INLA reflect the uncertainty conditional on the observed sample and the estimated inclusion probabilities \citep{williams2021a}. They do not propagate additional uncertainty arising from the adaptive sampling procedure itself, the estimation of sampling probabilities, Monte Carlo-estimated inclusion probabilities, or working model misspecification. In addition, the Leroux CAR prior may also smooth predictions too much near outbreak boundaries \citep{ward2023, gause2024}. Risk can change sharply between neighbouring locations at these boundaries, while the prior favours similar risks in nearby locations  \citep{duncan2020}.

There are several limitations in our study. Firstly, the simulation setting is simplified, with the number of individuals tested $n_{it}$ and population sizes set to the same, reasonably large, values across all locations (Appendix \ref{dgp}). Further simulations on a wider range of epidemic settings and varying test capacities and population sizes could provide a more comprehensive evaluation of the framework. Secondly, the sensitivity of the framework to the burn-in period, tuning parameters, and the choice of sampling budget in terms of both number $m$ of locations sampled and number $n_{it}$ of tests available has not been fully assessed. A short burn-in may leave early risk estimates unstable, whereas a long one delays adaptive sampling. However, the many years of historical sentinel sampling data (e.g. \citet{correa2016,gu2024}) could provide a readily available burn-in period. Finally, although IPW reduced estimation bias, this adjustment approach does not recover the full likelihood for the latent spatio-temporal process, because it does not specify a fully joint likelihood for both the risk process and the adaptive sampling mechanism \citep{lyddon2019}. It is instead a design-weighted loss used to construct a pseudo-posterior, making the model update explicitly design-aware \citep{bissiri2016}. Therefore, the IPW pseudo-posterior should be interpreted cautiously.  
Further work should test the framework across a wider range of settings and improve uncertainty calibration under adaptive sampling.

Overall, we have developed a Bayesian closed-loop adaptive surveillance framework that combines spatio-temporal risk prediction with flexible adaptive site allocation and design-aware IPW adjustment. The framework gives a unified approach for balancing outbreak detection, exploration, and inferential reliability under limited surveillance resources and demonstrates a promising avenue for improved infectious disease monitoring.

\PrintReferences
\label{lastpage}

\clearpage
\processdelayedfloats
\clearpage

\appendix{\textbf{Appendix}}

\makeatletter
\efloat@restorefloats %
\@removefromreset{figure}{section}
\@removefromreset{table}{section}
\SFB@appendixfalse
\makeatother

\setcounter{figure}{0}
\setcounter{table}{0}
\setcounter{section}{0}
\setcounter{secnumdepth}{3}

\renewcommand{\thefigure}{S\arabic{figure}}
\renewcommand{\thetable}{S\arabic{table}}
\renewcommand{\thesection}{A\arabic{section}}
\renewcommand{\thesubsection}{\thesection.\arabic{subsection}}

\section{Data generation process}
\label{dgp}

We generated surveillance data using a spatial stochastic susceptible--\allowbreak infectious--\allowbreak recovered--\allowbreak susceptible (SIRS) model to simulate epidemic dynamics and spatial transmission across the surveillance network. The simulation included $N=169$ locations arranged on a $13\times13$ grid over $T=60$ weeks, with a time-constant population of $N_i = 100,000$ individuals for each location $i$ and test capacity $n_{it} = 2,000$ for each $i$ and $t$. 
The data generation process is as follows:
\begin{enumerate}
    \item \textbf{Spatial SIRS dynamics}
    At each location $i$ and time step $k$, the population is divided into susceptible ($S_{ik}$), infectious ($I_{ik}$), and recovered ($R_{ik}$) compartments, with \(   S_{ik}+I_{ik}+R_{ik}=N_i \).
    Individuals transition from susceptible to infectious according to an infection hazard $\lambda_{ik}$, from infectious to recovered at rate $\gamma$, and from recovered to susceptible at immunity-loss rate $\omega$. The process is updated every time interval $\Delta$ using stochastic binomial transitions in a chain-binomial model, with compartment counts updated after each step. Therefore, independent draws conditional on the current spatial state are 
    \[
    B_{ik}\sim Binom(S_{ik},1-e^{-\lambda_{ik}\Delta}),\ D_{ik}\sim Binom(I_{ik},1-e^{-\gamma\Delta}),\ U_{ik}\sim Binom(R_{ik},1-e^{-\omega\Delta})
    \]
    and synchronous state updates are
    \[
    S_{i,k+1}=S_{ik} - B_{ik}+U_{ik},\ I_{i,k+1}=I_{ik} + B_{ik}-D_{ik},\ R_{i,k+1}=R_{ik}+D_{ik}-U_{ik}.
    \]
    
    \item \textbf{Spatial transmission and outbreak generation.} 
    The baseline infection hazard at location  $i$ and time $k$ is set to keep prevalence at a low endemic level. It includes local transmission $\lambda_{ik}^{(L)}$, transmission from neighbouring locations $\lambda_{ik}^{(N)}$, and infection from external sources $\lambda_{ik}^{(E)}$: 
    \[
    \lambda_{ik}=\lambda_{ik}^{(L)} +\lambda_{ik}^{(N)}+\lambda_{ik}^{(E)} = \beta_{ik}\left[(1-\psi)I_{ik}/N_i + \psi\sum_j\tilde A_{ij}I_{jk}/N_i\right] + h_i
    \]
    where $ \psi \in [0,1]$ is the neighbour mixing weight and $\beta_{ik}$ the transmission coefficient, which is modelled by Gaussian-shaped pulse functions as:
   \(\beta_{ik}=\beta_{i0} \left[1+a_i exp\left(-\frac{(x-\tau_i )^2
   }{2L_i^2 }\right) \right]\). Here, $\tau_i$ locates the centre of the transmission pulse, $L_i$ controls its width, and $a_i$ controls its amplitude, and $\beta_{i0}$ is determined by the baseline infection hazard. 

    \item \textbf{Surveillance observations. }
    Weekly true infection prevalence, $p_{it}$, was obtained by averaging the infectious proportion over the within-week simulation steps, $p_{it} = \frac{1}{M}\sum_{k \in K_t}\frac{I_{ik}}{N_i}$, where $M=7/\Delta$. Conditional on $p_{it}$ and the number tested $n_{it}$, observed positive counts were generated as
    \[
        Y_{it}\mid n_{it},p_{it}\sim\mathrm{Binomial}(n_{it},p_{it}).
    \]
    
    A location was classified as high risk when its underlying prevalence exceeded the threshold $\delta_0=0.1$, and an outbreak episode was defined as a consecutive sequence of high-risk weeks.

\end{enumerate}
\clearpage

\section{Figures}
\setcounter{figure}{0}

\begin{figure}[!htbp]
\centering
\InsertResultFigure{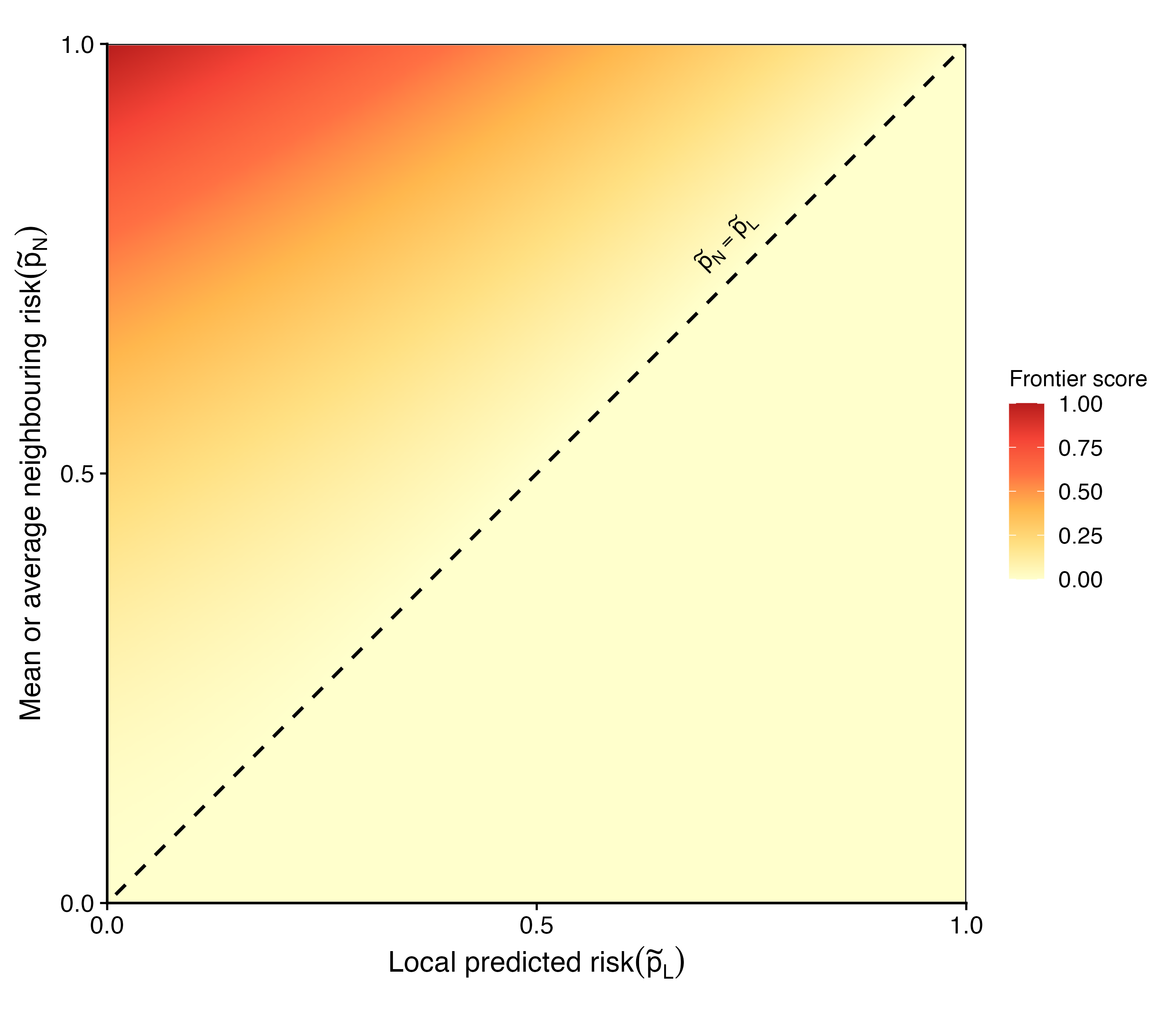}
\caption{\textbf{ Illustration of the frontier score.} The score is highest when local predicted risk is low and neighbouring risk is high. The dashed line indicates equal local and neighbouring risk $(\tilde p_N=\tilde p_L)$, below which the frontier score is zero. }
\label{fig:frontier}
\end{figure}

\begin{figure}[H]
\centering
\InsertResultFigure{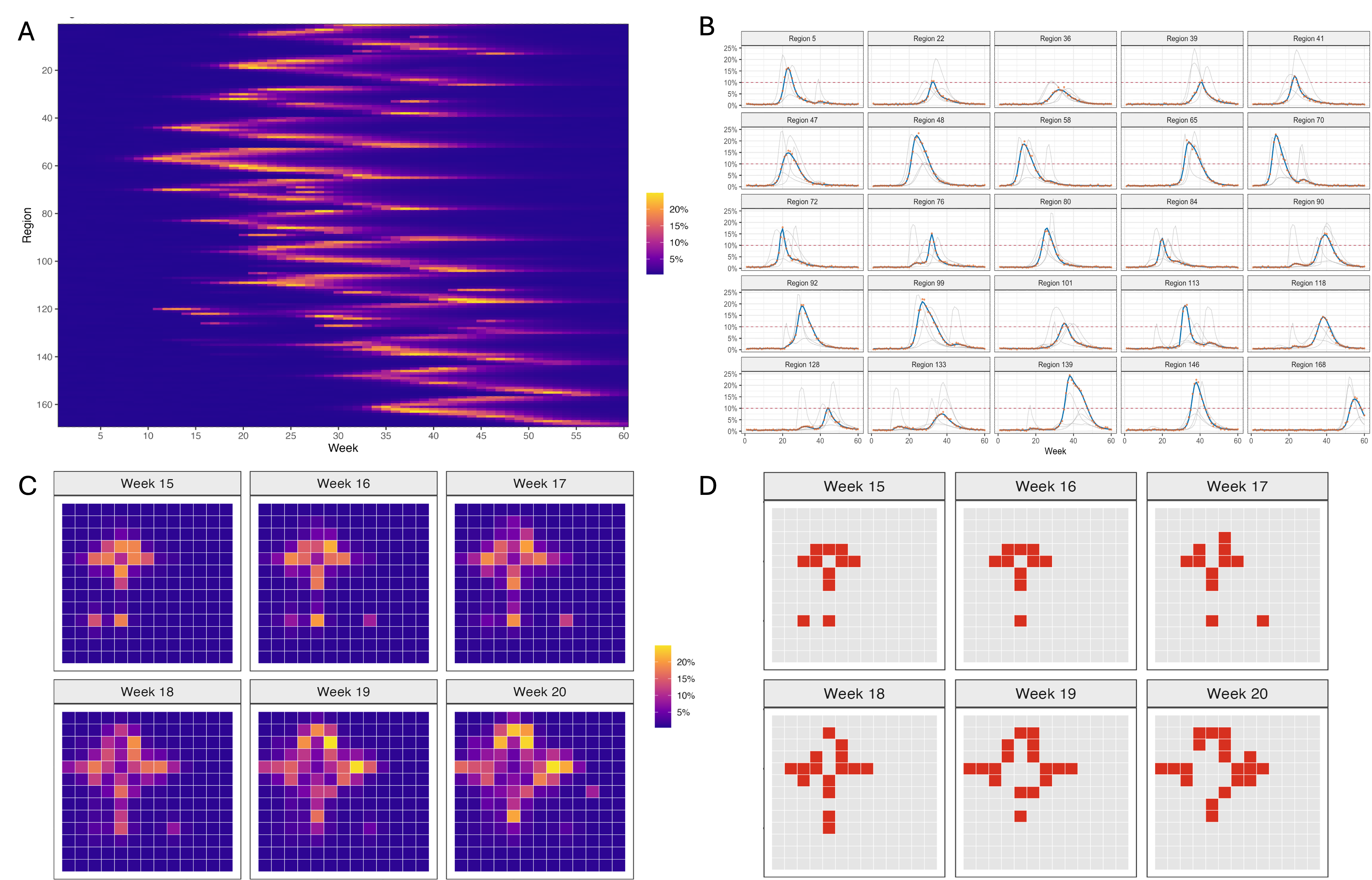}
\caption{\textbf{Illustration of the simulated spatiotemporal outbreak process.}
(A) Heatmap of the true prevalence \(p_{it}\) across 169 regions over 60 weeks, illustrating heterogeneous outbreak onset, duration, and peak prevalence across regions. (B) Temporal trajectories for selected regions and their neighbours. Blue curves denote the true prevalence in the central region, orange points the observed positivity \(y_{it}/n_{it}\), and grey curves the true trajectories of neighbouring regions. The red dashed line denotes the outbreak threshold \(\delta_0=0.10\). (C) Spatial maps of \(p_{it}\) from weeks 15 to 20. (D) Binary true outbreak states corresponding to panel C, where highlighted regions satisfy \(p_{it}>\delta_0\).}
\label{fig: heatmap}
\end{figure}

\clearpage

\begin{figure}[p]
\centering
\InsertResultFigure{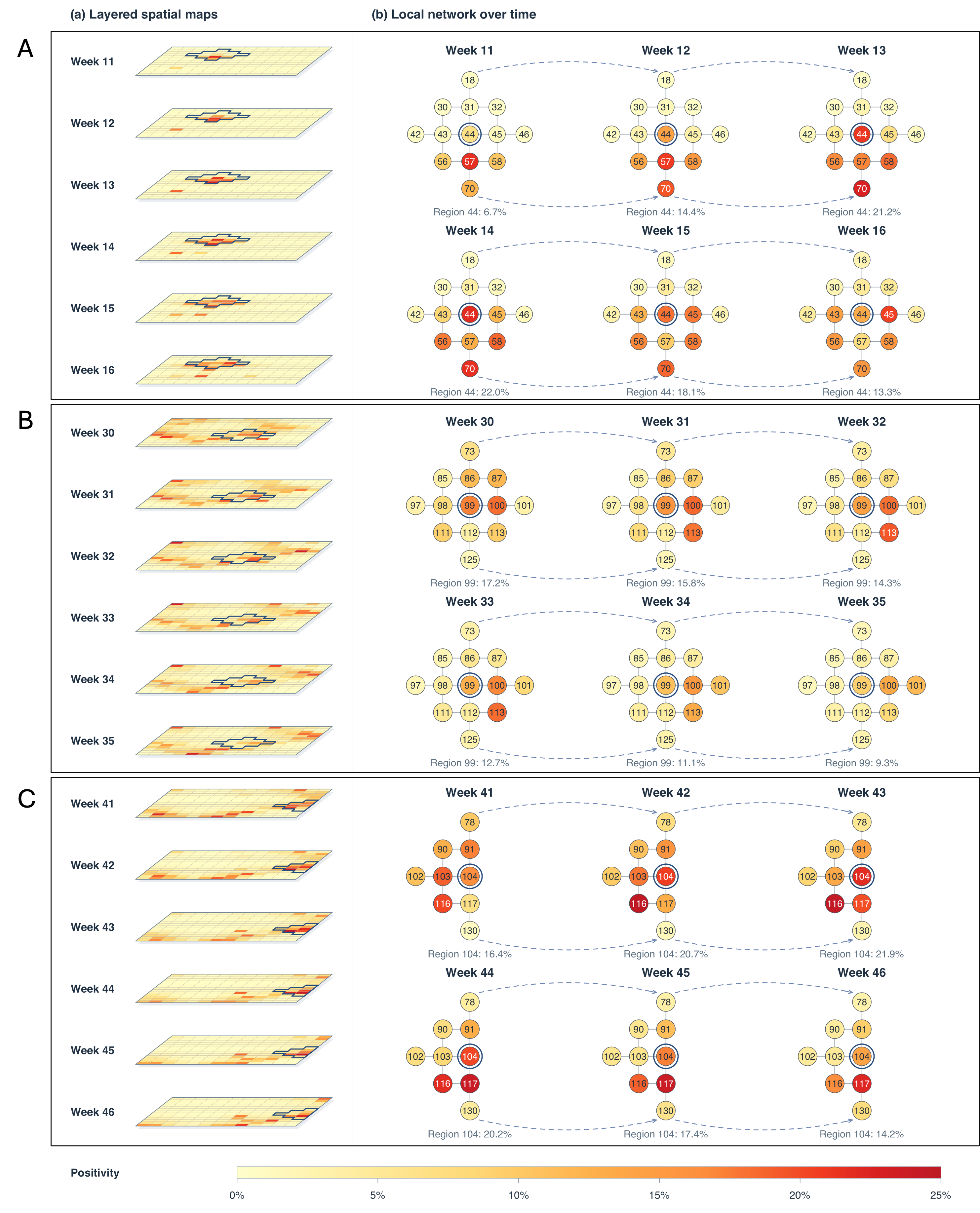}
\caption{Spatiotemporal patterns of prevalence across the surveillance network. 
Rows show local neighbourhoods centred on regions 44 (weeks 11–16)[A], 99 (weeks 30–35)[B], and 104 (weeks 41–46)[C]. (a) Layered weekly spatial maps retain the original grid positions, with blue outlines marking the areas within two grid steps of the central region displayed in panel (b). (b) The corresponding local networks followed over six consecutive weeks. Nodes represent regions, and blue rings identify the central regions. Solid lines denote four-neighbour spatial adjacency. Dashed arrows link the same regions across consecutive weeks.}
\label{fig: spatial_positivity}
\end{figure}

\clearpage

\clearpage

\clearpage

\section{tables}

\begingroup
\fontsize{10}{12}\selectfont
\setlength{\tabcolsep}{2pt}
\renewcommand{\arraystretch}{1.15}

\begin{table}[htbp]
\centering
\caption{Outbreak detection performance of different surveillance strategies with sampling budgets of 25 and 35 locations.}
\label{tab:budget_detection}
\small
\setlength{\tabcolsep}{4pt}
\renewcommand{\arraystretch}{1.12}
\resizebox{\textwidth}{!}{%
\begin{tabular}{llcccc}
\toprule
Budget & Strategy &
\shortstack{Detection\\rate}  &
Timely1 & Timely2 &
\shortstack{Mean delay\\(weeks)} \\
\midrule
35 & RP & 0.791(0.026) & 0.358(0.036) & 0.537(0.041) & 1.202(0.123) \\ 
   & RP-R & 0.784(0.031) & 0.293(0.038) & 0.496(0.045) & 1.365(0.137) \\ 
   & ANF & 0.819(0.029) & 0.400(0.039) & 0.596(0.038) & 1.043(0.125) \\ 
   & MNF & 0.813(0.028) & 0.399(0.039) & 0.592(0.034) & 1.037(0.119) \\ 
   & MNFC & 0.780(0.030) & 0.300(0.039) & 0.501(0.043) & 1.343(0.156) \\ 
   & URS & 0.649(0.046) & 0.191(0.034) & 0.351(0.044) & 1.777(0.227) \\ 
   & Sentinel & 0.205(0.014) & 0.189(0.015) & 0.205(0.014) & 0.076(0.050) \\ 
\midrule   
25 & RP & 0.655(0.038) & 0.217(0.030) & 0.363(0.036) & 1.752(0.219) \\ 
   & RP-R & 0.649(0.033) & 0.183(0.033) & 0.331(0.038) & 1.915(0.193) \\ 
   & ANF & 0.706(0.031) & 0.262(0.028) & 0.422(0.038) & 1.538(0.152) \\ 
   & MNF & 0.707(0.032) & 0.268(0.041) & 0.425(0.038) & 1.515(0.171) \\ 
   & MNFC & 0.672(0.038) & 0.210(0.035) & 0.369(0.040) & 1.719(0.187) \\ 
   & URS & 0.535(0.047) & 0.137(0.023) & 0.259(0.027) & 2.024(0.235) \\ 
   & Sentinel & 0.146(0.012) & 0.135(0.014) & 0.146(0.012) & 0.076(0.054) \\
\bottomrule   
   \end{tabular}%
}
\vspace{3pt}
\begin{minipage}{\textwidth}
\footnotesize
Values are means (standard deviations) across available simulation replicates.
Timely1 and Timely2 denote detection within the first one and two weeks, respectively.
\(\epsilon = 0.1\).
\end{minipage}
\end{table}
\clearpage

\begin{table}[htbp]
\centering
\caption{Risk estimation performance of different surveillance strategies with sampling budgets of 25 and 35 locations.}
\label{tab:budget_estimation}
\small
\setlength{\tabcolsep}{4pt}
\renewcommand{\arraystretch}{1.12}
\resizebox{\textwidth}{!}{%
\begin{tabular}{llcccccccc}
\toprule
 & & & & & & & \multicolumn{3}{c}{High-risk RMSE} \\
\cmidrule(lr){8-10}
Budget & Strategy & MAE & RMSE & Bias &
\shortstack{Coverage\\(95\% intervals)} &
Brier & Overall & Rising & Falling \\
\midrule
35 & RP & 0.011 & 0.027 & -0.0036 & 0.946 & 0.036 & 0.085 & 0.104 & 0.063 \\ 
    & RP-R & 0.011 & 0.027 & -0.0031 & 0.947 & 0.038 & 0.086 & 0.106 & 0.063 \\ 
    & ANF & 0.012 & 0.025 & 0.0006 & 0.970 & 0.033 & 0.074 & 0.090 & 0.057 \\ 
    & MNF & 0.012 & 0.025 & 0.0004 & 0.968 & 0.033 & 0.075 & 0.091 & 0.057 \\ 
    & MNFC & 0.012 & 0.026 & 0.0015 & 0.969 & 0.037 & 0.077 & 0.093 & 0.060 \\ 
    & URS & 0.018 & 0.043 & 0.0077 & 0.961 & 0.053 & 0.093 & 0.091 & 0.098 \\ 
    & Sentinel & 0.039 & 0.056 & 0.0334 & 0.989 & 0.061 & 0.059 & 0.067 & 0.050 \\ 
 \midrule
 25 & RP & 0.013 & 0.031 & -0.0045 & 0.933 & 0.046 & 0.098 & 0.116 & 0.077 \\ 
    & RP-R & 0.014 & 0.031 & -0.0041 & 0.936 & 0.047 & 0.098 & 0.117 & 0.076 \\ 
    & ANF & 0.014 & 0.029 & 0.0002 & 0.961 & 0.042 & 0.086 & 0.103 & 0.068 \\ 
    & MNF & 0.014 & 0.029 & 0.0005 & 0.962 & 0.041 & 0.086 & 0.102 & 0.068 \\ 
    & MNFC & 0.015 & 0.030 & 0.0022 & 0.966 & 0.044 & 0.085 & 0.101 & 0.068 \\ 
    & URS & 0.021 & 0.047 & 0.0097 & 0.958 & 0.062 & 0.096 & 0.098 & 0.098 \\ 
    & Sentinel & 0.046 & 0.061 & 0.0392 & 0.989 & 0.068 & 0.062 & 0.070 & 0.052 \\ 
 \bottomrule
\end{tabular}%
}
\vspace{3pt}
\begin{minipage}{\textwidth}
\footnotesize
Values are means across available simulation replicates. $\epsilon = 0.1$. High-risk RMSE is calculated over location–week observations for which the true prevalence exceeds the outbreak threshold \((p^{\mathrm{true}}_{it}>\delta_0)\).
\end{minipage}
\end{table}

\clearpage

\begin{table}[htbp]
\centering
\caption{Comparison of unweighted and IPW-updated adaptive surveillance frameworks under different exploration levels.}
\label{tab:ipw_comparison}
\renewcommand{\arraystretch}{0.95}
\resizebox{\textwidth}{!}{%
\begin{tabular}{lllccccccccc}
\toprule
$\epsilon$ &
Strategy &
Update &
\shortstack{Detection\\rate } &
Timely1 &
Timely2  &
\shortstack{Mean delay\\(weeks)} &
MAE &
RMSE &
Bias &
\shortstack{Coverage95} &
Brier \\
\midrule
 0.1 & RP & None & 0.791 & 0.358 & 0.537 & 1.202 & 0.0111 & 0.0269 & -0.0036 & 0.946 & 0.036 \\ 
       & RP & IPW & 0.785 & 0.345 & 0.520 & 1.270 & 0.0114 & 0.0277 & -0.0041 & 0.937 & 0.038 \\ 
       & RP-R & None & 0.784 & 0.293 & 0.496 & 1.365 & 0.0114 & 0.0272 & -0.0031 & 0.947 & 0.038 \\ 
       & RP-R & IPW & 0.775 & 0.279 & 0.479 & 1.420 & 0.0117 & 0.0277 & -0.0030 & 0.943 & 0.040 \\ 
       & ANF & None & 0.819 & 0.400 & 0.596 & 1.043 & 0.0116 & 0.0251 & 0.0006 & 0.970 & 0.033 \\ 
       & ANF & IPW & 0.821 & 0.403 & 0.594 & 1.066 & 0.0117 & 0.0254 & 0.0004 & 0.966 & 0.034 \\ 
       & MNF & None & 0.813 & 0.399 & 0.592 & 1.037 & 0.0116 & 0.0252 & 0.0004 & 0.968 & 0.033 \\ 
       & MNF & IPW & 0.824 & 0.411 & 0.603 & 1.012 & 0.0116 & 0.0249 & 0.0006 & 0.967 & 0.033 \\ 
       & MNFC & None & 0.780 & 0.300 & 0.501 & 1.343 & 0.0124 & 0.0265 & 0.0015 & 0.969 & 0.037 \\ 
       & MNFC & IPW & 0.787 & 0.302 & 0.510 & 1.343 & 0.0128 & 0.0269 & 0.0022 & 0.968 & 0.038 \\ 
       & URS & None & 0.649 & 0.191 & 0.351 & 1.777 & 0.0181 & 0.0425 & 0.0077 & 0.961 & 0.053 \\ 
       & URS & IPW & 0.655 & 0.198 & 0.358 & 1.760 & 0.0183 & 0.0429 & 0.0080 & 0.961 & 0.054 \\ 
   \hline
   0.2 & RP & None & 0.795 & 0.363 & 0.534 & 1.194 & 0.0111 & 0.0268 & -0.0035 & 0.946 & 0.036 \\ 
       & RP & IPW & 0.783 & 0.344 & 0.521 & 1.244 & 0.0113 & 0.0275 & -0.0039 & 0.938 & 0.038 \\ 
       & RP-R & None & 0.788 & 0.280 & 0.478 & 1.430 & 0.0114 & 0.0273 & -0.0027 & 0.950 & 0.039 \\ 
       & RP-R & IPW & 0.782 & 0.272 & 0.470 & 1.461 & 0.0117 & 0.0274 & -0.0022 & 0.949 & 0.040 \\ 
       & ANF & None & 0.831 & 0.415 & 0.618 & 0.986 & 0.0123 & 0.0253 & 0.0028 & 0.976 & 0.032 \\ 
       & ANF & IPW & 0.826 & 0.418 & 0.611 & 0.999 & 0.0123 & 0.0255 & 0.0025 & 0.972 & 0.033 \\ 
       & MNF & None & 0.827 & 0.426 & 0.621 & 0.946 & 0.0123 & 0.0251 & 0.0027 & 0.975 & 0.032 \\ 
       & MNF & IPW & 0.826 & 0.420 & 0.619 & 0.955 & 0.0122 & 0.0252 & 0.0025 & 0.972 & 0.032 \\ 
       & MNFC & None & 0.787 & 0.312 & 0.511 & 1.334 & 0.0125 & 0.0268 & 0.0018 & 0.970 & 0.037 \\ 
       & MNFC & IPW & 0.781 & 0.289 & 0.491 & 1.387 & 0.0127 & 0.0270 & 0.0017 & 0.967 & 0.039 \\ 
       & URS & None & 0.650 & 0.191 & 0.351 & 1.783 & 0.0181 & 0.0426 & 0.0077 & 0.961 & 0.053 \\ 
       & URS & IPW & 0.649 & 0.191 & 0.351 & 1.777 & 0.0181 & 0.0425 & 0.0077 & 0.961 & 0.053 \\ 
       \bottomrule
\end{tabular}%
}

\vspace{0.5em}

\begin{minipage}{\textwidth}
\footnotesize
Values are means across available simulation replicates. None denotes the unweighted model update and IPW denotes the inverse-probability-weighted update. The weekly sampling budget is 35 locations.
\end{minipage}
\end{table}
\endgroup
\clearpage

\end{document}